\documentclass{JHEP}
\usepackage[mathscr]{eucal}
\usepackage{graphicx}
\newcommand{\td}{\textup{d}}  
\def\Pcm#1{{\mathcal{#1}}}
\newcommand{\del}{\partial}
\def\eqref#1{(\ref{#1})}
\def\er#1{eqn.\eqref{#1}}
\def\nn{\nonumber}
\title{Field Definitions, Spectrum and Universality in Effective String Theories}
\author{N.~D.~Hari Dass \\ Hayama Center for Advanced Studies, Hayama, Kanagawa 240-0193, Japan \\ \email{hari@soken.ac.jp}}
\author{Peter Matlock \\ Department of Electrophysics, National Chiao Tung University, Hsinchu, Taiwan \\ \email{pwm@mail.nctu.edu.tw}}
\abstract{
It is shown, by explicit calculation, that the third-order terms in
inverse string length in the spectrum of the effective string theories
of Polchinski and Strominger are also the same as in Nambu-Goto
theory, in addition to the universal L\"uscher terms.  While the
Nambu-Goto theory is inconsistent outside the critical dimension, the
Polchinski-Strominger theory is by construction consistent for any
space-time dimension.  In the analysis of the spectrum, care is taken
not to use any field redefinition, as it is felt that this has the
potential to obscure important points. Nevertheless, as field
redefinition is an important tool and the definition of the field
should be made precise, a careful analysis of the choice of field
definition leading to the terms in the action is also
presented. Further, it is shown how a choice of field definition can
be made in a systematic way at higher orders.  To this end the
transformation of measure involved is calculated, in the context of
effective string theory, and thereby a quantum evaluation made of
equivalence of theories related by a field redefinition.  It is found
that there are interesting possibilities resulting from a redefinition
of fluctuation field.
}
\preprint{Sokendai-HKTK/061203}

\begin{document}

\section{Introduction}
\label{intro}
Fundamental string theories can only be consistently quantised in the
so-called critical dimension which is $D=26$ for bosonic and $D=10$
for supersymmetric theories.  On the other hand string-like defects or
solitons occur in a wide variety of physical circumstances, the most
well-known being vortices in superfluids, the Nielsen-Olesen vortices
of quantum field theories, vortices in Bose-Einstein condensates and
QCD strings.  These objects do clearly exist in dimensions other than
the previously mentioned critical dimensions.  The challenge then is
to find consistent quantum descriptions of such objects without 
restriction on the dimension.

Polchinski and Strominger (PS) \cite{PS} indeed showed how to do this.
Their proposal is in spirit very close to that of chiral perturbation
theory \cite{anant}, which is an effective description of QCD at low
energies. While requiring the symmetries of QCD to be maintained, it
is otherwise unconstrained by requirements like polynomial lagrangians
and renormalisability.  Likewise PS advocated including all possible
terms in the action that preserve the constraint and symmetry
structure of string theories.  The action terms they propose are not
polynomial and in fact can become singular for certain string
configurations. However, understood as terms in an effective action,
they are to be used in a \emph{long-string} vacuum, which 
allows perturbation in the small parameter $R^{-1}$ where $2\pi R$ is the
length of the (closed) string. Then the dominant term in the action is the 
usual quadratic action. 
In their construction, PS dropped terms in the
action which are proportional to equations of motion and constraints,
to appropriate orders in $1/R$. The quadratic action is equivalent to the
well known Nambu-Goto (NG) theory when suitable constraints are imposed. The spectrum
of the NG theory is exactly calculable as shown first by Arvis \cite{Arv}.

In the present work, we extend the original results of Polchinski and
Strominger to the next order in $1/R$. In doing this, we find that the
action need not be modified beyond what PS originally proposed
\cite{PS}.  We work first within the PS scheme and prove that there
are no corrections to the PS action at order $R^{-3}$. Using this
result, we find that there are no subsequent further corrections to
the spectrum, which thus still does not deviate from that of
Nambu-Goto theory.  Of course, this comparison is made whilst noting
that the effective string theory construction is valid in any
dimension.  In this analysis of the spectrum, (and in our earlier
presentation in \cite{orig}), we have stayed entirely within the PS
formulation and we carefully avoid the use of any additional
ingredients such as field redefinitions.

Field redefinitions bring with them a number of new issues like
associated changes in measures and intrinsic arbitrariness; after our
analysis of the absence of corrections to the spectrum, we thus turn
to the definition of the theory itself, in the sense of a definition
of the field. We first carefully define and then investigate the
effects of field redefinitions, when considering further higher
corrections to the action.

With regard to the field definition, the PS prescription involves a
particular choice in which terms proportional to the leading-order
equation of motion do not appear in the action. While this may always
be chosen, we show how other choices may be made, and cast some light
on the question of which terms are important in the action in a
general expansion, concluding with the possibility that even the PS
term is unnecessary, to certain orders, provided the transformation
laws leaving the action invariant are chosen suitably. In general,
there is a delicate interplay between the transformation law and the
field definition; only the combined specification of the field
variables, the action and the transformation laws defines a theory.

The plan of the paper is as follows. In section \ref{leadorder} we
briefly review the original PS scheme which is valid to order
$R^{-2}$.  We then prove in section \ref{sec_absence}, in very general
terms, the absence of additional terms in the action which are of
order $R^{-3}$.  This is crucial in establishing our results in
\ref{sec_high}.

Using this proof we carry out in section \ref{sec_high} an analysis of
the spectrum to higher orders, where we show the absence of both
order-$R^{-2}$ terms and corrections to order-$R^{-3}$ terms of the
spectrum of NG theory. These results have already been presented in
\cite{orig}, which forms the basis for sections \ref{sec_absence} and
\ref{sec_high}.  This result for the spectrum has also been shown in
\cite{Drum}\footnote{This unpublished work came to our attention while preparing
the manuscript \cite{orig}.}, although using somewhat different
methods, and making use of field redefinition. Also, the absence of
$R^{-3}$ terms in the action was merely asserted in \cite{Drum}
without systematic proof. The authors have had some debate
\cite{drumresp,dmresp} with the author of that paper, and we shall
point out differences and issues as appropriate, particularly in
sections \ref{fieldredef} and \ref{morefred}.

Field definitions and equivalence of theories in the effective-string
context are investigated in sections \ref{equivft} and
\ref{fieldredef}.  We finally conclude in section \ref{conclude} with
some remarks, observations, and ideas for further investigation.


\section{Leading-order analysis}
\label{leadorder}
Here, we review the analysis given by Polchinski and Strominger \cite{PS}.
They begin with the action
\begin{eqnarray}
\label{action2}
S &=& \frac{1}{4\pi} \int \td\tau^+ \td\tau^- \bigg\{
 \frac{1}{a^2} \del_+ X^\mu \del_- X_\mu \mbox{}+\beta \frac{\del_+^2 X\cdot\del_- X \del_+ X\cdot\del_-^2 X}{(\del_+X\cdot\del_-X)^2} 
+\Pcm{O}(R^{-3})
\bigg\}
.\end{eqnarray}
They show that this action is invariant, i.e $\delta S < \Pcm{O}(R^{-2})$, under the modified conformal transformations
\begin{equation}
\label{modtrans}
\delta_- X^\mu = \epsilon^-(\tau^-)\del_- X^\mu - \frac{\beta a^2}{2}\del_-^2 \epsilon^-(\tau^-)
\frac{\del_+ X^\mu}{\del_+X\cdot\del_-X}
,\end{equation}
(and another; $\delta_+X$ with $+$ and $-$ interchanged). Actually it turns out that if we
do not truncate the action in eqn(\ref{action2}) to $\Pcm{O}(R^{-2})$, $\delta S < \Pcm{O}(R^{-3})$. It can also be shown that the PS transformation law also closes to this higher order. Thus 
both the PS action and their transformation law can be considered consistently  to include
order-$R^{-3}$ terms. 
The full equation of motion (EOM), $E^\mu=0$, from the untruncated action is
\begin{eqnarray}
\label{modEOM}
E^\mu &=& -\frac{1}{2\pi a^2}\del_{+-}X^\mu
+\frac{\beta}{4\pi}\bigg[
       \del_+^2\big\{
          \frac{\del_-X^\mu(\del_-^2X\cdot\del_+X)}{L^2}\big\} \nn\\
&&+2\del_+\big\{\frac{\del_-X^\mu(\del_+^2X\cdot\del_-X)(\del_-^2X\cdot\del_+X)}{L^3}\big\} 
\nn\\
       &&-\del_-\big\{\frac{\del_+^2X^\mu(\del_-^2X\cdot\del_+X)}{L^2}\big\} 
+\{+\leftrightarrow -\}\bigg]
,\end{eqnarray}
where we have used the notation $L=\del_+X\cdot\del_-X$.
It is easy to see that
\begin{equation}
\label{classbg}
 X^\mu_{\textup{cl}} = e^\mu_+R\tau^+ + e^\mu_- R \tau^- 
;\end{equation}
where $e_-^2=e_+^2=0$ and $e_+\cdot e_- = -1/2$ satisfies the full EOM.
Fluctuations around the classical solution are denoted by $Y^\mu$, so that
\begin{equation}
X^\mu = X^\mu_{\textup{cl}}+Y^\mu
.\end{equation}
When fluctuations are considered around the \emph{long-string vacuum} characterised by
large R, it becomes meaningful to use $R^{-1}$ as an expansion parameter. Then the leading
order EOM is $\del_{+-}X^\mu$.
The untruncated transformation law
leads to the energy momentum tensor (which agrees with eqn(11) of \cite{PS}
to the relevant order)
\begin{eqnarray}
\label{fullT__}
T_{--}^{\textup{PS}} &=& -\frac{1}{2a^2}\del_-X\cdot\del_-X
+\frac{\beta}{2L^2}\big(
     L\del_-^2 L-(\del_- L)^2 \nn\\
&&+\del_-X\cdot\del_-X \del_+^2 X\cdot\del_-^2 X -\del_+ L \del_-X\cdot\del_-^2 X 
     \big)
\end{eqnarray}
where we have omitted terms proportional to the leading-order equation
of motion. This is justified because these terms are at most
proportional to $R^{-1}$ and from eqn(\ref{modEOM}) the leading order
EOM is itself of order-$R^{-2}$ at the most and the neglected terms
are therefore order-$R^{-3}$, whereas for the higher-order analysis to
be considered later, it suffices to keep upto order-$R^{-2}$ terms in
$T_{--}$.

Returning to the leading order-analysis of PS, the energy-momentum
tensor in terms of the fluctuation field is then
\begin{equation}
\label{T_2}
T_{--} = -\frac{R}{a^2}e_\cdot\del_-Y -\frac{1}{2a^2}\del_-Y\cdot\del_-Y
-\frac{\beta}{R}e_+\cdot\del_-^3 Y+\ldots
\end{equation}
with the OPE of
$T_{--}(\tau^-)T_{--}(0)$ given by 
\begin{equation}
\frac{D+12\beta}{2(\tau^-)^4}+\frac{2}{(\tau^-)^2}T_{--}
+\frac{1}{\tau^-}\del_-T_{--} 
+\Pcm{O}(R^{-1})
.\end{equation}

It should be noted that due to the 
$-\frac{R}{a^2}e_-\cdot\del_-Y$ term in $T_{--}$, in principle the order-$R^{-2}$
term in the $Y$-$Y$ propagator could contribute. 
It turns out that for the PS field definition it does not. 
When eqn(\ref{modEOM}) is restricted to terms linear in $Y^\mu$ we 
get an equation from which the two-point function can be computed;
\begin{equation}
\langle Y^\mu Y^\nu \rangle = -a^2\log(\tau^+\tau^-)\eta^{\mu\nu}
+2\frac{\beta a^4}{R^2}
e_-^{(\mu} e_+^{\nu)}\delta^2(\tau)
\end{equation}

Consequently the potential contribution to the central charge
$\frac{R^2}{a^4} e_-^\mu e_-^\nu \langle Y^\mu Y^\nu \rangle$
vanishes, as $e_-\cdot e_- =0$. This is not always true as can be
checked by redefining the $X^\mu$ field. Of course the total central
charge does not change. One must add the contribution $-26$ from the
ghosts, leading to the total central charge $D+12\beta-26$. Vanishing
of the conformal anomaly thus requires
\begin{equation}
\beta_c = -\frac{D-26}{12}
,\end{equation}
valid for any dimension $D$.

Using standard techniques the spectrum of this effective theory can be
worked out. PS have shown how to do this at the leading order.  We
briefly reproduce their results here in order to set the stage for the
rest of the paper. The Virasoro generators operate on the Fock space
basis provided by 
$\del_-Y^\mu = a\sum_{m=-\infty}^{\infty}\alpha_m^\mu e^{-im\tau^-}$ 
and are given by
\begin{eqnarray}
\label{virasoro}
L_n &=& \frac{R}{a}e_-\cdot\alpha_n
+\frac12\sum_{m=-\infty}^\infty :\alpha_{n-m}\cdot\alpha_m: 
+\frac{\beta_c}{2}\delta_{n}
-\frac{a\beta_c n^2}{R}e_+\cdot\alpha_n + \Pcm{O}(R^{-2})
.\end{eqnarray}

The quantum ground state is $|k,k;0\rangle$ which is also an
eigenstate of $\alpha_0^\mu$ and ${\tilde\alpha}_0^\mu$ with common
eigenvalue $ak^\mu$.  This state is annihilated by all $\alpha_n^\mu$
for positive-definite $n$.  The ground state momentum is
$p^\mu_{\textup{gnd}} = \frac{R}{2a^2}(e_+^\mu+e_-^\mu) + k^\mu$ while
the total rest energy is
\begin{equation}
\label{restenergy}
(-p^2)^{1/2} = \sqrt{
\left(\frac{R}{2a^2}\right)^2
-k^2-\frac{R}{a^2}(e_++e_-)\cdot k } 
.\end{equation}
The physical state conditions $L_0 = {\tilde{L}}_0 = 1$ fix $k$, so that
\begin{equation}
\label{leadingvir}
k^1 = 0,\qquad
k^2+\frac{R}{a^2}(e_++e_-)\cdot k = \frac{(2-\beta_c)}{a^2}
.\end{equation}
The first follows from the periodic boundary condition for the closed string
which gives $e_+^\mu-e_-^\mu = \delta^\mu_1$. Substituting the critical value
$\beta_c = (26-D)/12$ one arrives at
\begin{equation}
(-p^2)^{1/2} = \frac{R}{2a^2}\sqrt{1-\frac{D-2}{12}\left(\frac{2a}{R}\right)^2}
,\end{equation}
which is the precise analog of the result obtained by Arvis for open strings
\cite{Arv}. Expanding this and keeping
only the first correction, one obtains for the static potential
\begin{equation}
V(R) = \frac{R}{2a^2} -\frac{D-2}{12}\frac{1}{R}+\cdots
.\end{equation}


\section{Absence of additional terms at order $R^{-3}$}
\label{sec_absence}
It is of crucial importance for the arguments of this paper that the
next possible candidate term in the action is not $R^{-3}$ order.  PS
have stated without proof in \cite{PS} that the next such term is
actually of order $R^{-4}$. Drummond has stated, again without proof,
that further terms do not appear until order $R^{-6}$ \cite{Drum}.
However, as it is such a vital point, we give here the most general 
proof for absence of terms at order $R^{-3}$, which we had already
presented in \cite{orig}. 

We follow PS and construct actions that are $(1,1)$ in the na\"ive
sense; that is, the net number of $(+,-)$ indices is $(1,1)$.  We
include no terms proportional to the leading order constraints
$\del_\pm X\cdot \del_\pm X$ or to the leading order equations of
motion $\del_{+-} X^\mu$; otherwise they can be of arbitrary
form. Clearly such actions can be constructed out of skeletal forms of
the type
\begin{equation}
\label{skeleton}
\frac{X^{\mu_1}_{s_1,m_1}X^{\mu_2}_{s_2,m_2} \cdots X^{\mu_N}_{s_N,m_N}}{L^M}
\end{equation}
by contracting the Lorentz indices $\mu_1,\mu_2,...,\mu_N$ with the
help of \emph{invariant tensors} $\eta_{\mu\nu}$ and
$\epsilon_{\mu_1\mu_2..\mu_D}$. Let us consider the potentially
\emph{parity-violating} terms involving the Levi-Civita symbols
later. Here $X^\mu_{s,m}$ stands for $m$ derivatives of type $s=\pm$
acting on $X^\mu$.  The numbers $\{m_i\},M$ are adjusted to achieve
the (na\"ive) $(1,1)$ nature.

The PS lagrangian is not strictly a $(1,1)$ form as can be checked
explicitly. However the PS action, to the desired accuracy, is
invariant under the transformation laws of \er{modtrans}. It is
$(1,1)$ only in the na\"ive sense mentioned above.  The na\"ive
criterion is necessary but not sufficient, thus it suffices to prove
the absence of action terms that are $R^{-3}$ using this criterion.
In fact, it is desirable to have a formulation that is manifestly
covariant.  This will be presented elsewhere \cite{HDPM2}.

Only powers of $L$ have been used in the denominator to get a $(1,1)$
form. It may appear that any scalar in target space would have
sufficed. However, the action should not become \emph{singular} on any
fluctuation. Thus a scalar, say, of the type $\del_+^2 X\cdot\del_-X$
would not be permissible as it vanishes with $Y$. Whatever is in the
denominator must be of the form $\del_+X\cdot\del_-X+\cdots$; this can
always be expanded around the dominant $L$ term to produce forms as in
\er{skeleton}. A covariant formulation \cite{HDPM2} gives a natural
explanation for this as well as for the forms considered in
\er{skeleton}.

For now we will adhere closely to the PS prescription and discard all
\emph{irrelevant terms}, that is, terms proportional to the leading-order
equation of motion. With regard to the meaning and consistency of this procedure,
we return later in section \ref{cfd} where we explain that this
choice of irrelevant terms (i.e. none of them) in fact represents 
a particular choice of field definition. This should be compared with
\cite{drumresp} where it is suggested that the correct way of handling irrelevant
terms is to first eliminate them by a suitable redefinition of the field and then
work out the consequences for the action and transformation laws.

\subsection{Parity-conserving Sector}
\label{sec_pcs}
All those cases where the Lorentz contractions produce additional
factors of $L$ can be reduced to forms with lower $N$; we therefore
need not consider cases where the number of factors with higher
derivatives ($m\ge 2$) is smaller than the number with only single
derivatives. On the other hand cases with more higher-derivative
factors than single-derivative factors are less dominant. Thus for the
even-$N$ case considered first (taken as $2N$ from now onwards) we
need to consider the maximal case of exactly $N$ single-derivative
terms and $N$ terms with all possible higher derivatives.

Among the single-derivative terms, let $n_+$ be the number with
$+$-derivatives; then there are $N-n_+$ single derivative terms with
$-$-derivatives. Among the higher derivative terms let $p_+$ be the
number of terms with only $+$-derivatives, and likewise $p_-$. Let
$m_+$ be the total number of higher $+$-derivatives and $m_-$ the
corresponding number of higher $-$-derivatives.  As 
\begin{equation}
p_++p_- = N,\quad m_+\ge 2p_+, \quad m_-\ge 2p_-
,\end{equation}
it follows that $m_++m_-\ge 2N$,
\begin{equation}
 m_++n_+=m_-+N-n_+, \quad M = m_++n_+-1,
\end{equation} 
and subsequently that $2m_+\ge 3N - 2n_+$.

Now the leading-order behaviour of such a term is
$R^{N-2(m_++n_+-1)}$. On noting that $N+2-2n_+-2m_+ \le 2-2N$ we see
that for $N\ge 3$ the leading behaviour of the action is 
$R^{-4}$ at most\footnote{Drummond has observed \cite{drumresp} that when N is odd one
actually has $2m_++2n_+ \ge 3N +1$; then the leading behaviour is
actually $R^{-5}.$}. The case $N=2$ is precisely the PS action with $R^{-2}$
behaviour.  The dominant case among the subdominant class for $N=2$
(four factors) is where there are three factors with only higher
derivatives and one with a single derivative which we can take to be
of $+$-type without loss of generality. If $l_+$ denotes the total
number of $+$-derivatives among higher derivatives and likewise $l_-$,
we must have $l_--l_+=1$. As before, if $P_+$ denotes the number of
terms with only $+$-derivatives and likewise $p_-$, we have
$p_++p_-=3$ and then $l_+ \ge 2p_+$, $l_-\ge 2p_-$ and $l_++l_-\ge 6$.
These lead to $l_-^{min} =4,l_+^{min}=3$, giving $M=3$ and the
leading-order behaviour is then of order $R^{-5}$. For $N=1$ (two
factors) we can only have higher-derivative terms and it is easy to
see that the dominant term is $\del_+^2X\cdot\del_-^2X/L$, which in
the context of this analysis is equivalent to the PS action.

We have therefore proven that there is no possibility of an
order-$R^{-3}$ term, with the next-order terms potentially at order
$R^{-4}$ and $R^{-5}$.  Drummond has further refined \cite{drumresp}
our above analysis and shown that in fact these potential terms reduce
via partial differentiation to terms of order $R^{-6}$ and terms
proportional to the leading-order EOM, validating his original claim
in \cite{Drum}, at least in the parity-conserving case.

It is worth pointing out that though the demonstration \cite{drumresp}
of the absence of $R^{-4},R^{-5}$ order terms is certainly important,
it is not of much use until the PS transformation law is appropriately
modified as it closes only up to $R^{-4}$ order.  In general, giving
possible terms in the action without discussing the transformation
laws that would leave them invariant is incomplete. It could well be
that there are no transformation laws that leave some or all of them
invariant.

Finally, one may note that just as the absence of additional terms in
the action of order $R^{-3}$ does not automatically imply, as is
evident from both \cite{Drum} and our analysis , the absence of
$R^{-3}$ corrections to the NG spectrum, absence of $R^{-4}$ or
$R^{-5}$ terms in the action also does not translate immediately into
any statement about the even higher order corrections to the spectrum.

\subsection{Parity-violating sector}
\label{sec_pvs}
Finally we turn to parity-violating cases and first to the case where
there is an \emph{odd} number of $X$ fields present.  This can only
happen when $D$ is odd, say, $2n+1$.  The contraction must be between
$\epsilon_{\mu_1..\mu_{2n+1}}$ and an expression of the form
\begin{equation}
\del_+ X^{\mu_1}\del_- X^{\mu_2}
\del_+^2 X^{\mu_3}\del_-^2 X^{\mu_4}\ldots 
\del_+^{n+1} X^{\mu_{2n+1}}
.\end{equation}
The total number of $+$-derivatives is $n(n+1)/2+n+1$, while the total
number of $-$-derivatives is $n(n+1)/2$. The above expression
multiplied by $\del_-^{n+2}X\cdot\del_+X$ balances the $+,-$
derivatives (terms with $+$ and $-$ interchanged are also allowed).
This has to be divided by $(\del_+X\cdot\del_-X)^{n(n+1)/2+n+1}$,
producing a leading behaviour of $R^{3-n^2-3n-2}$ or $R^{-(n^2+3n-1)}$. 
Clearly, when $D \ge 5$ the dominant bahaviour is at most $R^{-9}$.
In $D=3$ one can have
\begin{equation}
\label{3dpv}
L^{-3}\epsilon_{\mu_1\mu_2\mu_3}\partial_+X^{\mu_1}\partial_-X^{\mu_2}\partial_+^2 X^{\mu_3}\partial_+X\cdot\partial_-^3 X
\end{equation}
which has potential $R^{-3}$ behaviour. 
By partial integration we can recast \eqref{3dpv} as
\begin{eqnarray}
&-&L^{-3}\epsilon_{\mu_1\mu_2\mu_3}\partial_-X^{\mu_2}\partial_-({\partial_+X^{\mu_1}\partial_+^2 
X^{\mu_3}})\partial_+ X\cdot\partial_-^2 X \nn\\
&-&L^{-3}\epsilon_{\mu_1\mu_2\mu_3}\partial_+ X^{\mu_1}\partial_-^2 X^{\mu_2}\partial_+^2 
X^{\mu_3}~\partial_+ X\cdot\partial_-^2 X  \\
&-&3L^{-4}\epsilon_{\mu_1\mu_2\mu_3}\partial_+ 
X^{\mu_1}\partial_- X^{\mu_2}\partial_+^2 X^{\mu_3}(\partial_-^2 X\cdot\partial_+X)^2 \nn
.\end{eqnarray}
The first line produces irrelevant terms, meaning terms proportional
to leading EOM, of order $R^{-3}$ and higher. We will discuss later
how these and other irrelevant terms are treated while discussing the
use of field redefinitions.  Both the remaining lines are of order
$R^{-4}$ and higher. It should be noted that this has been done
without recourse to fluctuation field and thus automatically obeys our
principle of $X$-uniformity introduced later in section
\ref{fluctyxuni}.

If we do use the $Y$ field, we may further recast the $R^{-4}$ terms as
\begin{equation}
\frac{8}{R^4}\epsilon_{\mu_1\mu_2\mu_3}e_+^{\mu_1}\partial_+^2Y^{\mu_3}e_+\cdot\partial_-^2Y\big\{\partial_-^2 Y^{\mu_2}
 - 6 e_-^{\mu_2}e_+\cdot\partial_-^2 Y\big\}
\end{equation}
These can again be reduced to irrelevant terms by partial integration,
and therefore $R^{-4}$ terms may also be eliminated, at the expense of
$X$-uniformity.  Nevertheless, the $R^{-5}$ terms remain and we see no
way to get rid of them, at least as of now.

In the $D=4$ parity-violating case 
\begin{equation}
L^{-2}\epsilon_{\mu_1\mu_2\mu_3\mu_4}\partial_+ X^{\mu_1}\partial_- X^{\mu_2}\partial_+^2 X^{\mu_3}
\partial_-^2 X^{\mu_4}
\end{equation}
the order-$R^{-2}$ term 
$\epsilon_{\mu_1\mu_2\mu_3\mu_4}e_+^{\mu_1}e_-^{\mu_2}\partial_+^2 Y^{\mu_3}\partial_-^2 Y^{\mu_4}$, 
can be eliminated by partial integration as it reduces to
\begin{equation}
\epsilon_{\mu_1\mu_2\mu_3\mu_4}e_+^{\mu_1}e_-^{\mu_2}\partial_{+-}Y^{\mu_3}\partial_{+-}Y^{\mu_4}
\end{equation}
due to the complete antisymmetry of the $\epsilon_{\mu_1\mu_2\mu_3\mu_4}$. 

We thus conclude that in the parity violating sector there are indeed
$R^{-4}$- and $R^{-5}$-order terms.  We return later to the issue of
how the irrelevant terms at order $R^{-3}$ should be handled in the
$D=4$ context also.


\section{Higher corrections to ground-state energy}
\label{sec_high}
From the expression for the ground-state momentum, it is clear that
all higher corrections are determined by
$A(k) =k^2+\frac{R}{a^2}(e_++e_-)\cdot k$ (\er{restenergy}) which was only
calculated to leading order in \er{leadingvir}. Thus an order-$R^{-n}$
correction to this would result in order-$R^{-n-1}$ and higher
corrections to the spectrum. As the original PS analysis gives $R^{-1}$ terms
in the spectrum, we need to investigate both $R^{-1}$
and $R^{-2}$ corrections to $A(k)$. As this quantity is just a sum of the $L_0$
and ${\tilde{L}}_0$ conditions, we need to calculate up to
order-$R^{-2}$ corrections to $L_0$ and ${\tilde L}_0$, or
equivalently to $T_{--}$.

As the transformation laws \eqref{modtrans} have a leading part linear
in $R$, additional terms in the action at order $R^{-3}$ would in
principle have induced $R^{-2}$ corrections to $T_{--}$. That would in
turn have changed the $R^{-3}$ terms in the spectrum. This is the
reason why the absence of such terms in the action needs to be
established so carefully. Absence of such terms also means that the
expression for $T_{--}$ in \er{fullT__} can be consistently expanded
to keep order-$R^{-2}$ terms. We give here the \emph{on-shell}
expression to the desired order;
\begin{eqnarray}
\label{highT}
&&T_{--}= -\frac{R}{a^2}e_-\cdot\del_-Y-\frac{1}{2a^2}\del_-Y
\cdot\del_-Y-\frac{\beta}{R}e_+\cdot\del_-^3Y \nn\\
&&-\frac{\beta}{R^2}\big[2(e_+\cdot\del_-^2Y)^2
+2e_+\cdot\del_-^3Y(e_+\cdot\del_-Y+e_-\cdot\del_+Y) \nn\\
&&+2e_-\cdot\del_-^2Ye_-\cdot\del_+^2Y+\del_+Y\cdot\del_-^3Y
\big]
.\end{eqnarray}

We see hence that $L_0$ and ${\tilde{L}}_0$ do not receive any
order-$R^{-1}$ correction.  At this point, the $T_{--}$ of \er{highT}
does not seem \emph{holomorphic} as there are $+$-derivative terms
occurring in $T_{--}$, while the Noether procedure necessarily gives a
$T_{--}$ which satisfies $\del_+ T_{--}=0$.  The resolution of this
apparent contradiction lies in the fact that the solution of the full
equation of motion \eqref{modEOM} can no longer be split into a sum of
holomorphic and antiholomorphic pieces.

Because of the absence of additional terms in the action with the
leading $R^{-3}$ behaviour, the equation of motion \eqref{modEOM} is
sensible inclusive of $R^{-3}$ terms. Now we expand this expression
and retain terms up to order $R^{-3}$;
\begin{eqnarray}
\label{R3EOM}
\frac{2}{a^2}\del_{+-}Y^\mu =
- 4\frac{\beta}{R^2}\del_+^2\del_-^2 Y^\mu 
&-&4\frac{\beta}{R^3}\bigg[
\del_+^2\big\{
              \del_-^2Y^\mu(e_+\cdot\del_-Y+e_-\cdot\del_+Y)
                \big\} \nn\\
        &+&\del_-^2\big\{
               \del_+^2Y^\mu(e_+\cdot\del_-Y+e_-\cdot\del_+Y)
                 \big\}  \nn\\
 &-&  e_+^\mu\del_-(\del_+^2\cdot\del_-^2Y)-e_-^\mu\del_+(\del_+^2Y\cdot\del_-^2Y)
\bigg]
.\end{eqnarray}
We can solve this equation iteratively by writing $Y^\mu = Y^\mu_0+Y^\mu_1$ 
where $Y^\mu_0$ is a solution of the leading order equation of motion.
Keeping terms only up to order $R^{-3}$ we obtain an 
expression which can be readily integrated to yield
\begin{eqnarray}
\frac{2}{a^2}\del_{-}Y_1^\mu &=& 4\frac{\beta}{R^3}
\big( e_+^\mu\del_+Y_0\cdot\del_-^3Y_0+e_-^\mu\del_+^2Y_0\cdot\del_-^2Y_0 \nn\\
&&-\del_-^2Y_0^\mu e_-\cdot\del_+^2Y_0-\del_+Y_0^\mu e_+\cdot\del_-^3Y_0 \big) 
.\end{eqnarray}
Examining \er{highT} one sees that to order $R^{-2}$ only the first term linear 
in $R$ contributes additional non-holomorphic terms which \emph{exactly} 
cancel the remaining non-holomorphic pieces.
This immediately leads to the \emph{manifestly holomorphic} representation 
of $T_{--}$ to order $R^{-2}$,
\begin{eqnarray}
T_{--} &=& -\frac{R}{a^2}e_-\cdot\del_-Y_0-\frac{1}{2a^2}\del_-Y_0\cdot
\del_-Y_0 -\frac{\beta}{R}e_+\cdot\del_-^3Y_0 \nn\\
&-&2\frac{\beta}{R^2}e_+\cdot\del_-^3Y_0e_+\cdot\del_-Y_0 
-2\frac{\beta}{R^2}(e_+\cdot\del_-^2Y_0)^2
,\end{eqnarray}
whence we obtain the Virasoro generators with higher-order corrections,
\begin{eqnarray}
\label{vircorr}
L_n &=& \frac{R}{a}e_-\cdot\alpha_n+\frac{1}{2}\sum_{m=-\infty}^\infty:\alpha_{n-m}\alpha_m:
+\frac{\beta_c}{2}\delta_{n} \nn\\
&&-\frac{a \beta_c n^2}{R}e_n 
-\frac{\beta_c a^2 n^2}{R^2}\sum_{m=-\infty}^\infty :e_{n-m}e_m: 
,\end{eqnarray}
where $e_n\equiv e_+ \cdot \alpha_n$.

Thus we have established that $L_0$ and ${\tilde{L}}_0$ have no
corrections at either $R^{-1}$ or $R^{-2}$ order. As mentioned
earlier, this means all the terms in the ground state energy and the
excited state energies, inclusive of the order-$R^{-3}$ term, are
identical to those in the Nambu-Goto theory.


\section{Fluctuation Field Y and X-uniformity}
\label{fluctyxuni}
We have made use of an expansion about a classical vacuum in our
arguments about the order of terms in the action, and further made use
of this fluctuation field to calculate the spectrum, along the lines
of PS \cite{PS}. Furthermore, we will below consider some issues
arising in field redefinitions, and we here tentatively put forth some
ideas involving the fluctuation field in this context.

With no loss of generality one can make what amounts to a functional
shift and set $X^\mu = X_{\textup{cl}}^\mu + Y^\mu$ where
$X_{\textup{cl}}^\mu = R(e_+^\mu\tau^++e_-^\mu\tau^-)$ which, as in
\eqref{classbg} is a solution of the equation of motion of the free
part of the action. Since all additional terms in the effective action
involve higher derivatives, $X_{\textup{cl}}^\mu$ continues to be a
solution of the full EOM, and $Y^\mu$ continues to have the
interpretation of a fluctuation around a classical background.

While carrying out field redefinitions of the $Y$-field, care should
be taken not to upset the $(0,0)$ nature of the $X$-field. As this property
is not explicit in the $Y$-formulation, it is very easy to upset.

Since the fluctuation field $Y$ derives from the $X$-field, one could
consider the following principle of ``$X$-uniformity'': \emph{All
expressions involving $Y^\mu$ and its derivatives, $e_{\pm}^\mu, R$
must be such they are derivable from expressions involving $X^\mu$ and
its derivatives only}.

Many field redefinitions of the $Y$-field will not be permissible if
this principle is adopted. In fact the field redefinitions found by
Drummond involving $R^{-2}$ and $R^{-3}$ terms of the
Polchinski-Strominger action are of a type that do not satisfy
this principle.

If this principle of $X$-uniformity is applied, the $(0,0)$ property
of $Y^\mu$ is never upset. In fact, without this principle, it seems
unclear how to define precisely the $(0,0)$ property for the
fluctuation field. Still, this does not necessarily mean that such a
definition is impossible in case $X$-uniformity is not imposed.

Further motivation for the imposition of $X$-uniformity is as
follows. When it is broken, the fluctuation field $Y$ and derived
spectrum thereof no longer have a direct interpretation in terms of
excitations about some background, since of course it is not possible
in such a case to write down the background in question.  Without such
an interpretation, the physical significance or meaning of the
fluctuations may be difficult to appreciate.

On the other hand, if one is given an effective theory describing
fluctuations, there is \emph{a priori} no clear reason why one should
not define the field differently, and indeed one could set out
with such an effective theory without even having seen the original 
$X$ field. 

\section{Equivalence of field theories}
\label{equivft}
As much of the discussion below will centre around the issue of
equivalence of field theories under a change of variables (what is
called field redefinition) we review here known facts and establish
some new points in the context of effective string theories.

As far as effective string theories are concerned two distinct
possibilities exist as to their interpretation as quantum
theories. One is to treat them as a tool to calculate the so-called
tree diagrams only.  In the early days of effective chiral symmetric
field theories (of pions, for example) this was the attitude taken; in
this case, equivalence of field theories would only mean equivalence
as in classical field theories, as explained below in section
\ref{ecf}.

Now, it is well known in chiral perturbation theory that such a
limited approach would have missed many essential features like
universal logarthmic behaviour and would have made a discussion of
analytic properties of scattering amplitudes beyond perview. The more
modern approach is to make the calculation of even loop amplitudes
meaningful and the lack of renormalisability is handled through a
larger set of arbitrary parameters. If we interpret the effective
string theories this way, all features of a full quantum field theory
are to be considered.

\subsection{Classical field theories}
\label{ecf}
Equivalence in classical field theories is easily proved; see for
example \cite{kamefuchi} for an explicit demonstration for the case of
\emph{point transformations}, namely, cases where the redefinition does
not involve time derivatives. At a classical level this can be
extended also to more general redefinitions.  This classical
equivalance means that for such things as the currents and
energy-momentum tensor, either using the lagrangian formalism on the
original lagrangian and then transforming the fields to a new set
gives the same results as first transforming the fields and then
applying the formalism.  The point is that the change of variables and
the lagrangian formalism commute, and a field redefinition can thus be
done at will to simplify the procedure without further thought.

When the redefinition in question is an \emph{infinitesimal} one it is
possible, classically, to eliminate terms in the action that are
proportional to the leading order (in an appropriate sense) EOM.
Equivalently, it can be stated that two classical actions differing
from each other by EOM terms in this sense are physically equivalent.
Therefore the PS prescription \cite{PS} of dropping all EOM terms as
\emph{irrelevant} is certainly justified at a classical level.  The
important question is to what extent this procedure is also valid
quantum-mechanically.  We are able to make some progress on this issue
at least within the path-integral formulation. We argue, somewhat
formally, that up to order $R^{-3}$ classical equivalence also implies
quantum equivalence, but beyond that things are uncertain.  This means
that our original proof of the absence of order-$R^{-3}$ corrections
to the spectrum is valid and complete even when considering quantum
equivalence. It also means that the
claim of absence of order-$R^{-4}$ and -$R^{-5}$ corrections in the
action made by Drummond in \cite{Drum} is certainly true classically
but its validity quantum-mechanically is not certain.

\subsection{Canonical formulation of quantum field theories}
\label{cqft}
Proofs of equivalence in the canonical, or operator, formalism of
quantum field theories are not completely straightforward, the main
obstacle being the non-commutativity of various operator expressions;
this has been very clearly analysed in
\cite{kamefuchi,apfeldorf}. Although in that paper they do manage to
prove the quantum equivalence for point transformations, such a proof
for field redefinitions that go beyond point transformations does not
seem to exist. In the context of effective string theories the field
redefinitions considered are certainly not of the point-transformation
type.

We shall therefore analyse this issue within the path-integral
formalism. There is of course an intimate connection (and usually
assumed equivalence) between the operator and path-integral
formulations of quantum field theories. Though there are very
important structural and conceptual differences, the final results for
physically relevant issues are expected to be the same. This
equivalence between canonical and path-integral formulations is subtle
and delicate as shown long ago by Lee and Yang \cite{leeyang}.

\subsection{Path-integral formulation}
\label{pif}
In the path-integral formulation, given a particular field definition,
there are three important issues: the action; the transformation laws
leaving the action invariant; the invariance of the measure under said
transformation laws. For non-linear transformation laws the
tranformation of the measure can of course be quite involved.

In the case of the PS transformation law, it is easy to see that the
na\"ive measure ${\cal D}X$ is indeed not invariant. We calculate now
the change in the na\"ive measure under the PS transformation
law. Consider the relation between the untransformed field $Y^\mu$ and
the (infinitesimally) transformed field ${Y^\prime}^\mu$;
\begin{eqnarray}
{Y^\prime}^\mu &=& Y^\mu+R\epsilon^-~e_-^\mu+\epsilon^-\partial_-~Y^\mu+\frac{\beta~a^2}{R}e_+^\mu\partial_-^2\epsilon \nn\\
&&+\frac{\beta~a^2}{R^2}\partial_-^2\epsilon^-\{\partial_+Y^\mu+2e_+^\mu(e_-\cdot\partial_+Y+e_+\cdot\partial_-Y)\} \nn\\
&&+\frac{2\beta~a^2}{R^3}\partial_-^2\epsilon^-e_+^\mu\{2(e_+\cdot\partial_-Y+e_-\cdot\partial_+Y)^2+\partial_+Y\cdot\partial_-Y\}
+\cdots
.\end{eqnarray}
Denoting the na\"ive measure by ${\cal D}Y$, its change under the above transformation is
${\cal D}Y^\prime = {\cal J}{\cal D}Y$,
where ${\cal J}$ is the determinant of 
${\cal J}^\mu_\nu(\xi,\xi^\prime)= \frac{\delta {Y^\prime}^\mu(\xi^\prime)}{\delta Y^\nu(\xi)}$
given by
\begin{eqnarray}
&&{\cal J}^\mu_\nu(\xi,\xi^\prime) =\delta^\mu_\nu\delta^2(\xi-\xi^\prime)+\delta^\mu_\nu\epsilon^-\partial_-\delta^2(\xi-\xi^\prime)
+\frac{\beta~a^2}{R^2}\delta^\mu_\nu\partial_-^2 \epsilon^-\partial_+\delta^2(\xi-\xi^\prime) \nn\\
&&+\frac{2\beta~a^2}{R^2}\partial_-^2\epsilon^-e_+^\mu (e_{+\nu}\partial_-\delta^2+ e_{-\nu}\partial_+\delta^2)
+ \frac{2\beta~a^2}{R^3}\partial_-^2\epsilon^-\partial_+Y^\mu(e_{+\nu}\partial_-\delta^2+e_{-\nu}\partial_+\delta^2) \nn\\
&&+\frac{2\beta~a^2}{R^3}\partial_-^2\epsilon^-\delta_\nu^\mu\partial_+\delta^2(e_-\cdot\partial_+Y+e_+\cdot\partial_-Y) \nn\\
&&+\frac{2\beta~a^2}{R^3}\partial_-^2\epsilon^-e_+^\mu(4(e_{+\nu}\partial_-\delta^2+e_{-\nu}\partial_+\delta^2)
(e_-\cdot\partial_+Y+e_+\cdot\partial_-Y) \nn\\
&&\qquad\qquad+(\partial_-Y_\nu\partial_+\delta^2+\partial_+Y_\nu\partial_-\delta^2))+\cdots
.\end{eqnarray}
The appearance of $\delta^2(\xi-\xi^\prime)$ and its derivatives makes
the evaluation of ${\cal J}$ very delicate and a proper regularisation
scheme, like for example some form of $\zeta$-function regularisation,
is needed to do this carefully.  There are two possible fates for these
singular terms.

One possibility is that the singular terms from the measure could
cancel against singular non-covariant expressions arising on a careful
use of Feynman rules in the apparently covariant path integral. This
is what was shown in \cite{leeyang}. The same situation
shows up in perturbative quantum gravity also \cite{gdep}.

The other possibility is that there are singular terms in the measure
which do not vanish or cancel in the above fashion, leaving
non-trivial factors to be taken into account. The only relevant
changes are contained in the {\em $Y$-dependent} part of ${\cal J}$
and these come from those parts of the transformation law having at
least quadratic terms in the fluctuation field $Y^{\mu}$.

We conclude that at least formally, by which we mean that of course a
careful regularised definition of the path integral in principle
should be taken into account, one can say that the variation of the
measure under the PS transformation law is order-$R^{-3}$ and hence
does not affect our work in sections \ref{sec_high} where only
variations that are of order $R^{-2}$ are relevant.  Nevertheless, the
important point evident from the above expression is that when any
analysis is carried out for order-$R^{-4}$ terms and beyond, this
could become an important issue and must be handled with care.

To the orders relevant for our original proof, the invariant measure is
therefore just the naive measure. At higher orders this is no longer
true and the issue of the correct measure has to be properly dealt
with.

\subsection{Jacobian for field redefinitions}
\label{fredefj}
When a generic field redefinition (change of variables in the path
integral) is made, the new action is obtained from the old by simple
substitution. The new transformation laws are what are induced by the
redefinition. Unless the field redefinition is invertible it is not
possible to work out the induced transformation laws. For infintesimal
changes in the fields, this can always be done.

Of course, this is not the whole picture; the change in the measure
for path integration consequent to the change in variables should also
be taken into account.  Simple counting arguments of the type
developed above in section \ref{sec_absence} and improved upon in
\cite{drumresp} show that the most general field redefinitions have
the structure
\begin{equation}
\Delta Y^\mu = c_1(R) [Y] + c_2(R) [YY] + c_3(R) [YYY]+\cdots
\end{equation}
where $[Y],[YY],\dots$ symbolically denote terms that are linear,
quadratic, etc.~in the $Y$-field. Further it can be shown that the
coefficients $c_n(R)$ can {\em at most} be of order $R^{-(n+1)}$. The
dominant field redefinition in fact has the form
\begin{eqnarray}
{Y^\mu}^\prime = Y^\mu&+& \frac{\alpha}{R^2}\partial_{+-} Y^\mu  
+\frac{\beta}{R^3}\partial_{+-}Y^\mu (e_+\cdot\partial_-Y)
+\cdots \nn\\
&+&\frac{\gamma}{R^4}\partial_{+-}Y^\mu(\partial_+Y\cdot\partial_-Y)
+\cdots
\end{eqnarray}
The jacobian for this transformation can again be formally computed as
before. Again we encounter singular expressions and careful
regularisation is required to make further progress. The non-trivial
part of the jacobian, by which we mean the $Y$-dependent terms, can
only come from terms that are {\em cubic} in the $Y$-field in the
field redefinition and those are of order $R^{-4}$. These are not
relevant for the proof we have presented in section \ref{sec_absence}
and in \cite{orig}.

We conclude this section with two points.  Firstly, as far as the
proof of the absence of $R^{-3}$ corrections to the Nambu-Goto
spectrum is concerned, quantum equivalence is the same as classical
equivalence.

Secondly, if a careful evaluation of these jacobians leads to local
terms, one may still be able to use Drummond's results \cite{drumresp} for
absence of terms to order $R^{-5}$ to show that classical equivalence
implies quantum equivalence up to this higher order. A rough argument
in support of this could be constructed by making field redefinitions
and performing partial integration as done by Drummond to eliminate
all terms up to order $R^{-6}$. An inspection of the field
redefinitions carried out in \cite{drumresp} reveals that at least
formally the resultant jacobian also does not get any contributions to
order $R^{-5}$. Nevertheless, this is only a sketch and we can have
confidence in it only after it is carefully established.  Until then,
quantum equivalence to orders $R^{-4}$ and beyond should be taken as
unproven.


\section{Field Definitions and Redefinitions}
\label{afr}
\label{fieldredef}

Our analysis of the spectrum, as we mentioned in the introduction,
does not involve a field redefinition.  Of course, one can make the
casual statement that irrelevant terms can be removed by a field
redefinition, and thus be led to think that such a redefinition has
been made when such terms are discarded. In fact, it is more precise
to say that we have through this procedure made a certain \emph{choice
of field definition}. as opposed to a \emph{field redefinition}.  This
is an important distinction which we now explain fully.

\subsection{Choice of field definition}
\label{cfd}
The PS procedure can be stated algorithmically as follows: Firstly,
write down all possible $(1,1)$ terms (in the sense used by PS);
Secondly, discard all terms proportional to the leading order
constraints and their derivatives; Finally, use integration by parts
to relate equivalent terms.

At this point one will have terms with and without `mixed
derivatives', terms sporting mixed derivatives being what we have
called irrelevant.  The PS prescription then is to discard all
irrelevant terms and \emph{find transformation laws} that leave the
relevant terms in the action invariant.

This is still not uniquely specified as not all the relevant terms are
independent and some can be related to others through integration by
parts and additional irrelevant terms. The unambiguous method is to
first express all the relevant terms in terms of a particular choice
of a minimal set (this choice in itself being arbitrary) and
additional irrelevant terms. Next, one chooses a subset of the
irrelevant terms and simply drops all the rest. Finally, one finds
transformation laws that leave this combination of relevant and
irrelevant terms invariant (modulo issues of quantum equivalence
discussed earlier).

Now a particular choice of mixed derivative terms amounts to a
\emph{choice of field definition}. A different choice of the
irrelevant terms amounts to yet another choice of field
parametrisation. As long as the conditions for equivalence under field
redefinitions hold, one can go from one parametrisation to another
with the help of a \emph{field redefinition}. The transformation laws
in the new definition can be worked out as induced by the field
redefinition (this is somewhat more than mere substitution).

Let us illustrate this with an example. Take the partial set of terms
at order $R^{-2}$ to be
\begin{eqnarray}
\label{sample}
&&\alpha \frac {\partial_+^2 X\cdot\partial_-^2 X}{L}+\beta \frac {\partial_+^2 X\cdot\partial_- X
\partial_-^2 X\cdot\partial_+ X}{L^2}\nonumber\\
&+&\delta \frac {\partial_{+-}X\cdot\partial_{+-}X}{L}
+\eta \frac{\partial_{+-}X\cdot\partial_+ X
\partial_{+-}X\cdot\partial_-X}{L^2}
\end{eqnarray}
In \cite{Drum} it is claimed that the two relevant actions here are
equivalent modulo total derivatives. This is not true and this has
some bearing on the issues discussed; using the identity
\begin{eqnarray}
\frac{\partial_+^2 X\cdot\partial_-^2 X}{L} &=& 
\frac {\partial_+^2 X\cdot\partial_- X
\partial_-^2 X\cdot\partial_+ X}{L^2}\nonumber\\
&+&\frac {\partial_{+-}X\cdot\partial_{+-}X}{L}
- \frac{\partial_{+-}X\cdot\partial_+ X
\partial_{+-}X\cdot\partial_-X}{L^2}\nonumber\\
&+&\partial_-\big(\frac{\partial_+^2X\cdot\partial_-X}{L}\big)
-\partial_+\big(\frac{\partial_{+-}X\cdot\partial_-X}{L}\big)
\end{eqnarray}
\eqref{sample} can be rewritten as
\begin{eqnarray}
&&\beta^\prime \frac {\partial_+^2 X\cdot\partial_- X
\partial_-^2 X\cdot\partial_+ X}{L^2} 
+\delta^\prime \frac {\partial_{+-}X\cdot\partial_{+-}X}{L}
+\eta^\prime \frac{\partial_{+-}X\cdot\partial_+ X
\partial_{+-}X\cdot\partial_-X}{L^2} 
\end{eqnarray}
with $\beta^\prime = \beta + \alpha$,$\delta^\prime = \delta + \alpha
$ and $\eta^\prime = \eta - \alpha$. Now the choice $\delta^\prime
=0,\eta^\prime = 0$ yields one field definition, say, $X^\prime$ with
the effective action
\begin{equation}
\beta^\prime \frac {\partial_+^2 X^\prime\cdot\partial_- X^\prime
\partial_-^2 X^\prime\cdot\partial_+ X^\prime}{{L^\prime}^2}
\end{equation}
while the choice $\beta^\prime = \alpha$,$\delta^\prime = \alpha$ and
$\eta^\prime = -\alpha$ gives another field definition, say, $X^*$
with the effective action
\begin{equation}
\alpha \frac{\partial_+^2 X^*\cdot\partial_-^2 X^*}{L^*}  
\end{equation}
Thus specific choices for coefficients of the mixed derivative terms
merely pick out specific parametrisations and have nothing to do with
redefinitions. Having chosen a particular parametrisation, one has to
work out the transformation laws leaving the action invariant, the
consequent stress tensors, and so on.

In this particular example, though the two forms of relevant terms are
related by a non-trivial field redefinition, the transformation laws
for the two cases are identical. In fact there are families of field
redefinitions that do not induce any additional terms in the
transformation laws; for examples see section \ref{redefnt}. Of
course, in generic cases field redefinitions induce additional terms
in the transformation laws.

We conclude this section with two somewhat subtle points about such
redefinitions. Firstly, care has to be taken to carry through this
procedure consistently to all orders. In particular, it should be
noticed that with field redefinitions of order $\epsilon$ there is a
residual error of at least order $\epsilon^2$ which must be included in the
analysis of the most general effective action to be carried out to the
next higher order.  The other subtle but important point is that a
redefinition eliminating one set of irrelevant terms may bring another
set through the transformation of the measure.

\section{Field redefinitions and irrelevant terms}
\label{morefred}

With regard to field redefinitions, we have examined issues of
measures and jacobians in section \ref{fredefj} above.  Whereas
clearly such matters cannot simply be asserted not to affect the
calculations, in light of our analysis, it appears that indeed these
may not be issues of consequence to the orders required in proving the
absence of corrections to the spectra in the present work and in
\cite{Drum}.

As an example of the sort of issue which can arise, let us consider
how a particular field redefinition is used in \cite{Drum}. In that
paper, a field redefinition\footnote{eqns.(2.17-2.19)} is used to
bring the more complicated Lagrangian for the fluctuation field $Y$ of
the PS action to a simple form. In fact, the redefinition of $Y$
reduces the PS action modulo the terms of order higher than $R^{-3}$
to the particular \emph{free action}
\begin{equation}
\label{simplefree}
{\cal L}^{(0)} = \frac{1}{4\pi a^2} \partial_+{\tilde Y}\cdot\partial_-{\tilde Y}
.\end{equation}
Now, looking at the action \eqref{simplefree}, the casual reader would 
have expected the standard stress tensor
\begin{equation}
\label{trivial}
-\frac{1}{2a^2}\partial_-{\tilde Y}\cdot\partial_-{\tilde Y}
,\end{equation} 
but instead one is confronted with a different and non-trivial energy
momentum tensor in \cite{Drum}. That is, the energy-momentum tensor,
if obtained from the old energy-momentum tensor via substiutution of
the redefined field $\tilde{Y}$, is non-trivial, and of course does
not coincide with \eqref{trivial}.

The issue at hand is how to make the correct `choice' of energy-momentum
tensor, between \er{trivial} and the non-trivial energy momentum
tensor obtained in \cite{Drum} by substitution of the field
redefinition.  Clearly this question is not spurious, as it involves a
redefinition of the field, and we in fact resolve this issue in
section \ref{rmf} below.

It is indeed not enough, as mentioned in \cite{drumresp}, for the
theory to remain conformal with critical central charge and
non-trivial energy momentum tensor. Under field redefinitions central
charge certainly cannot change, and neither can the conformal nature
of a theory; this is somewhat beside the point, as the issue is really
whether or not there are nontrivial effects on the spectrum.

Although it may appear \cite{drumresp} that our procedure of
iteratively solving field equations is equivalent to the procedure of
\cite{Drum} involving field redefinition, the following comparisons
can be made, showing that in fact there are significant differences.

In our approach, based on the iterative solution to the EOM, we have
no need to rework either the action or the transformation laws.  In
the approach of \cite{Drum}, both of these have to be done
\footnote{In
\cite{Drum}, the induced transformation law was not computed. We
compute it in section \ref{redefnt}.}.
  In addition to computation of
the action to obtain the two-point function relevant for the OPE
calculation, the method also requires the computation of the
transformation law to know whether the energy-momentum tensor obtained
by substitution is the canonical one or not.

In our method, as we explain in section \ref{cfd}, we make use of no
field redefinition. In consequence, there are no issues of quantum
equivalence. In this respect, differences between the two procedures
would be more pronounced at higher order.  Indeed, iterative solutions
to field equations are possible (in principle) to any order, but as is
mentioned in \cite{Drum}, certain subtleties of the method involving
field redefinition (at least in the present context of effective
strings) seem to be restricted to order $R^{-3}$.

In general solving the full EOM iteratively involves non-locality.  It
is a fortuitous circumstance here (perhaps because of 2-d) that it is
quasi-local. Field redefinitions are typically local (not necessarily
point transformations) by contrast, with nonlocal field redefinitions
being a largely uninvestigated and difficult subject.

All one could have concluded had one designed a field redefinition
based on the iterative solution to the full equations of motion is
that ${\tilde Y}$ would be solution of the EOM of \er{simplefree}, but
of course that does not guarantee that the action is given by
\er{simplefree}. There are many actions (including the quadratic part
of the terms in PS) whose EOM is satisfied by such a redefined
field. Significantly, they generically have different two-point
functions and consequently different OPEs. Such a procedure also does not guarantee
that the transformation law is the standard one associated with free
actions. One would have had to recompute both these, though the
latter was not done in \cite{Drum,drumresp}. In the case of \cite{Drum},
as we mentioned above at the beginning of section \ref{morefred}, while
the induced action turns out to be free, the induced transformation
law turns out to be non-trivial. Our method here does not require the
consideration of these issues.

Before further examining the nature of the irrelevant terms, we pause
here to note that in our analysis of possible terms in section
\ref{sec_absence} it took some care to decide whether a given term
vanishes (by partial integration), or alternately can be reduced to an
irrelevant (proportional to the EOM) term. In fact, this makes no
difference for our calculations, given the way we have handled the
irrelevant terms. Again making a comparison, using the method of
\cite{Drum,drumresp} this would make a difference, as what might have
been thought to be an irrelevant term, necessitating a field
redefinition and the attendant induced transformation law, could
actually be shown to vanish if it turns out to be a total derivative.
In our opinion, this motivates the use of the method in the present
paper where the field definition, as explained in section
\ref{fieldredef}, is made clear from the outset and the inclusion of
various irrelevant terms is merely a choice of parametrisation.

\subsection{Treating the irrelevant terms}
\label{rmf}

In investigating the possible $R^{-3}$ corrections to the spectrum, we
have dropped any order-$R^{-4}$ terms in the action.  It has been
suggested by Drummond \cite{drumresp} that this is inconsistent, and
the following explanation is offered.  It is shown in \cite{Drum}
that, in the field definition chosen in that paper, the PS action is
of order $R^{-4}$ and subsequently claimed that neglecting these
$R^{-4}$ terms can only be done after taking due account of the
changes in the transformation law induced by the field redefinition
that allowed the reduction of the PS action to $O(R^{-4})$. From this
point of view, our treatment of the parity violating terms where we
simply drop the $R^{-2}$ and $R^{-3}$ terms as being reducible to
mixed derivative terms must appear to be incorrect.

In fact, this explanation is overly complex, and the consistency and
correctness of our procedure is easily seen as follows. Summarising
once more, what we have done is drop all the irrelevant terms, and
find the symmetry variations for what is retained. This means that, to
the order we were originally interested in, the only relevant action
terms are the PS term and the free action; the PS transformation law
leaves them invariant, actually to order $R^{-3}$. Noting this,
nothing more need be done.

As we have explained, our treatment merely amounts to a specific
choice of field parametrisation and not to a redefinition.  We feel
that the distinction we made between these in section \ref{cfd} is an
important one.  In contrast to this straightforward procedure, the
alternative proposed in \cite{drumresp} is that one should
first have determined the transformation laws that would leave
invariant the free and PS terms along with the irrelevant terms in the
parity violation action, then carried out the field redefinition that
would remove the irrelevant terms, and finally computed the
modifications to the transformation laws.  We now show through
explicit calculation that these extra ingredients are unnecessary.

For this purpose consider a generic irrelevant term
\begin{equation}
{\cal L}_{\textup{irr}} = {\cal F}^\mu \partial_{+-}X_\mu
\end{equation}
where ${\cal F}^\mu$ can be any general expression constructed out of
derivatives of $X^\mu$. In particular it can also contain additional
mixed derivative terms.  Now we wish to modify the transformation law
so that it leaves ${\cal L}_{\textup{free}}+{\cal L}_{\textup{irr}}$
invariant.  That this can always be done follows from the fact that
the variation of the EOM is proportional to the EOM (in a functional
sense) which is just another way of saying the EOM is covariant under
symmetry variations.  Now we evaluate the variation of ${\cal
L}_{\textup{irr}}$ under
\begin{equation}
\delta^{(0)}X^\mu = \epsilon^-\partial_- X^\mu
,\end{equation}
yielding
\begin{eqnarray}
\delta^{(0)}{\cal L}_{\textup{irr}} &=& \delta^{(0)}{\cal F}^\mu\partial_{+-}X_\mu
      +{\cal F}^\mu\delta^{(0)}(\partial_{+-}X_\mu)\nonumber\\
&=& 
\delta^{(0)}{\cal F}^\mu\partial_{+-}X_\mu-\epsilon^-\partial_-{\cal F}^\mu\partial_{+-}X_\mu)
,\end{eqnarray}
where we have dropped total derivative terms as usual. Therefore the
modification to the transformation law under which 
${\cal L}_{\textup{free}}+{\cal L}_{\textup{irr}}$ will be 
invariant is
\begin{equation}
\delta^\prime X^\mu = 2\pi a^2(\delta^{(0)}{\cal F}^\mu-\epsilon^-\partial_-{\cal F}^\mu)
.\end{equation}
On the other hand the field redefinition needed to remove the irrelevant term is
\begin{equation}
\Delta X^\mu = -2\pi a^2 {\cal F}^\mu
.\end{equation}
This induces the additional terms
\begin{eqnarray}
{\tilde\delta}X^\mu&=&\delta^{(0)}\Delta X^\mu - \epsilon^-\partial_-(\Delta X^\mu)\nonumber\\
&=& 
- 2\pi a^2(\delta^{(0)}{\cal F}^\mu-\epsilon^-\partial_-{\cal F}^\mu)
\end{eqnarray}
and this contribution exactly cancels the modification demanded by the irrelevant term.

Generalisation of these arguments to include an arbitrary number of
relevant terms in the action is straightforward as long as one keeps
track of the order of terms carefully.

We conclude that the procedure suggested in \cite{drumresp} is
equivalent to simply dropping all irrelevant terms and working with
the transformation laws that leave only the relevant terms invariant,
which is exactly what we have done in the first sections of the
present work. As we have described, this is in fact a particular field 
definition. Thereafter dropping the irrelevant parity-violating
terms one is left with just the free action and PS terms as relevant;
as the PS transformation laws expanded to the next higher order left
them invariant, this exercise serves as a check on our methodology in
sections \ref{sec_absence} and \ref{sec_high}.

At this stage it may occur to the reader that following such a
procedure, the entire PS action itself might be dropped. Although it
has been suggested in \cite{drumresp} that this indicates there is
something amiss with our methodology, it is in fact a valid possibility
which we instead take seriously in the following section.


\subsection{Relevant part of the PS action}
\label{dps}
 
In the previous section we discussed our dropping of $\Pcm{O}(R^{-4})$
terms.  Naturally, one is tempted to see how far this procedure can be
carried, and to determine which terms must be retained and which may
be discarded.  Indeed, it did not go unnoticed \cite{drumresp} that
using our methods, one could be led to discard the PS term in the
action in its entirety.

At first this seems surprising, and an argument against it could
begin with the assertion that the PS action has left its trace in
terms of a non-trivial energy momentum tensor and a non-trivial
transformation law, somehow remembering the parameters (i.e. $\beta$)
of the irrelevant terms. Also, one could argue, it was the PS term 
that rendered the effective theories valid in all dimensions.

Following this putative argument, our irrelevant terms also could in
principle have left their nontrivial traces, but these are simply not
there in our way of handling things.  Consider, though, that we have
explicitly and quite generally proved in the previous section that the
field redefinitions exactly compensate any modifications to the
transformation laws brought forth by the irrelevant terms: How can
these two apparently contradicting situations be reconciled?

Before we go on we wish to categorically state that the $R^{-4}$-order
terms are totally irrelevant for our analysis in sections
\ref{sec_absence} and \ref{sec_high}.  This is also true for
Drummond's analysis in \cite{Drum}. The simple reason for this is that
the variation of such terms under the PS transformation law can
\emph{at most} be of order $R^{-3}$ and consequently their
contribution to the Virasoro generators can also be at most $R^{-3}$.
The above-mentioned imprints in the transformation law must all be
from $R^{-2}$ or $R^{-3}$ terms.

Now, let us go on and examine the case of PS action more carefully.  If
we had simply dropped the irrelevant terms from the PS action we would
have ended up only with (modulo $R^{-4}$ and higher order terms)
\begin{equation}
\frac{1}{2\pi a^2}\int \partial_+Y\cdot\partial_- Y
\end{equation}
Now according to our prescription we should just work with this
relevant action and the transformation laws leaving it invariant.
Somewhat surprisingly, it turns out that not just the standard
\begin {equation}
\delta_-^{(0)} Y^\mu = \epsilon^- \partial_-Y^\mu
.\end{equation}
but {\em at least} the entire two-parameter family
\begin{eqnarray}
\label{biggertrans}
{\bar\delta}^{(0)}_- Y^\mu &=& \epsilon^- \partial_- Y^\mu
 + \epsilon^- R~e_-^\mu \\
&+&\frac{\beta^\prime a^2}{R} \partial_-^2\epsilon^-~e_+^\mu
+\frac{2\beta^\prime a^2}{R^2}\partial_-^2\epsilon^- e_+^\mu e_+\cdot\partial_- Y \nn
\end{eqnarray}
leaves the above action invariant (exactly). Also, they constitute the
conformal group (for fixed $R$ and $\beta'$) as can be verified by
\begin{equation}
[\delta_-(\epsilon^-_1),\delta_-(\epsilon^-_2)] =  \delta_-(\epsilon^-_{12});~~~\epsilon^-_{12} = \epsilon^-_1\partial_-\epsilon^-_2 - \epsilon^-_2\partial_-\epsilon^-_1
\end{equation}
It should be emphasised that at this stage the parameter
$\beta^\prime$ has nothing to do with $\beta$.

The resulting canonical energy-momentum tensor is indeed the same with
$\beta$ replaced by $\beta^\prime$ as eqn(19) of \cite{orig} or
eqn(2.20) of \cite{Drum} (after correcting a sign error). The central
charge depends on the free parameter $\beta^\prime$ and can be
adjusted for consistency in all dimensions.

On the other hand, if we had included the irrelevant terms used
in \cite{Drum}, the transformation law that would have left the above
combination of relevant and irrelevant terms invariant would have been
modified to
\begin{equation} 
\label{total}
\delta_-^{tot} = {\bar\delta}_-^{(0)}Y^\mu+
\frac{\beta a^2}{R} \partial_-^2\epsilon^-~e_+^\mu
+\frac{2\beta a^2}{R^2}\partial_-^2\epsilon^- e_+^\mu e_+\cdot\partial_- Y
\end{equation}

If now one had done a field redefinition to remove the irrelevant
terms there would be additional terms induced in the above
equation. As shown above their effect would be to cancel the
$\beta$-dependent terms and just leave \er{biggertrans}.

If one computes the transformation law to which the PS-transformation
law would have been modified by the field redefinition, one finds it
to be just \er{biggertrans} with $\beta^\prime$ replaced by $\beta$,
modulo some harmless terms arising from the ambiguities in field
redefinitions.

The question now is: How did $\beta^\prime$ get to be replaced by
$\beta$ which is a parameter belonging entirely to the irrelevant
terms?  The resolution is the following. According to the above
construction what would leave the combination of relevant and
irrelevant terms of this example would be \er{total}, but this is
clearly not the PS-transformation law expanded to suitable orders.
Nevertheless in the scheme used in \cite{Drum}, this would have been
taken to be the PS-transformation suitably expanded. This is
consistent with \er{total} \emph{if and only if $\beta^\prime$ had
been equated to $\beta$}.

It seems then that, contrary to folklore, a free bosonic string theory
with an adjustable central charge can be made consistent provided the
conformal transformation law is chosen appropriately.  This can be
viewed as an alternative approach to the spectrum of free strings that
appears consistent in all dimensions.  How far such an approach to
string theory can be extended further is currently under investigation
\cite{freebos}.
\subsection{Ambiguities in field redefinitions}
\label{redefnt}
All field redefinitions are \emph{ambiguous} up to terms of the type
\begin{equation}
\Delta X^\mu = N^\mu, \qquad N\cdot E = 0
\end{equation}
where $E^\mu$ is the EOM.

Not all field redefinitions induce changes in transformation laws. 
Some examples are:
\begin{eqnarray}
\Delta_1 X^\mu &=& \frac{\partial_{+-}X^\mu}{L} \nn \\
\Delta_2 X^\mu &=& \frac{(\partial_+X^\mu\partial_{+-}X\cdot\partial_-X+\partial_-X^\mu
\partial_{+-}X\cdot\partial_+X}{L^2} \nn\\
\Delta_3 X^\mu &=& \frac{(\partial_+X^\mu\partial_{+-}X\cdot\partial_-X-\partial_-X^\mu
\partial_{+-}X\cdot\partial_+X}{L^2} \nn
\end{eqnarray}
It should be noted that $\Delta_3 X^\mu$ is of the form of an
ambiguity. It is so even when $L^{-2}$ is replaced by any power of
$L$, but the transformation law is unaffected only for $L^{-2}$.
Only the choice $L^{-2}$ maintains the $(0,0)$ character of $X^\mu$.


\section{OPE and Virasoro algebra to higher order}
\label{sec_opevho}
In the first part of this paper, sections \ref{sec_absence} and
\ref{sec_high}, we consistently expanded the PS action \cite{PS} to
order $R^{-3}$, and the PS stress tensor to order $R^{-2}$.  This
analysis also implies that it is enough to expand the transformation
law given by PS to include order-$R^{-2}$ terms. Incidentally, the
transformation law is closed (satisfying the Virasoro algebra) to
order $R^{-3}$; only at order $R^{-4}$ do the PS transformations fail
to close.

Here we also point out that the variation of the \emph{entire} PS
action, eqn (1) of \cite{orig}, without truncating to any order, under
the \emph{entire} PS variation, eqn(2) of \cite{orig}, again without
any truncations, has only $\beta^2$ terms and these are of order
$R^{-4}$.  Furthermore, the untruncated PS transformation has the
closure property of Virasoro algebra also to order $R^{-4}$ as is very
easily verified. The terms that spoil closure are again $\beta^2$
terms.  Also, the form of the higher corrections do not obviously affect
the central charge. Taken together these guarantee that there are no issues with
either the OPE of stress tensors or of the validity of the Virasoro
algebra generated by their moments to the order to which we extended
the PS results.  The fact that our higher order energy-momentum tensor
is conserved is another consistency check.


\section{Conclusions and comments}
\label{conclude}

Not only is the Polchinski-Strominger action \cite{PS} the unique
effective first-order action for a consistent conformal theory of long
strings, but as we have carefully shown in sections \ref{sec_absence}
and \ref{sec_high} it is essentially unique up to and including terms
of third order in the inverse string length. The only remaining
freedom in the action is the use of irrelevant terms, as for example
the other `equivalent' form of the PS term, what we have called the
other relevant term, which we mentioned in section \ref{sec_absence},
and which does not alter our results for the spectrum. As we have
discussed in section \ref{fieldredef}, this is an example of choice
of field definition, and the possibilities become more numerous as one
considers higher-order actions.

Furthermore, the spectrum is found to coincide with that of the
Nambu-Goto theory, including third order terms.  This universality
explains why comparisons between potentials and excited state energies
in lattice computations \cite{LW,pushan,kuti,caselle,kuti2} and
Nambu-Goto theory have been favourable in the past even beyond the
universal L\"uscher term \cite{pushandass1,pushandass2} (in the case of the ground
state energy), despite the inconsistency of the Nambu-Goto string
outside the critical dimension.

Again with regard to the above statements about subsequent
higher-order terms in the action, it has now been proven carefully
\cite{drumresp} that the action does not contain relevant terms all
the way up to and including order-$R^{-5}$ terms, when consideration
is restricted to the parity-conserving sector. This is significant,
and one would like to ask what consequence this has for corrections to
the spectrum. An obvious path for further work is to determine when
corrections to the Nambu-Goto spectrum do in fact occur. In the
parity-violating sector, we have shown in section \ref{sec_pvs} that
correction terms begin at order $R^{-5}$ and it would also be
interesting to consider the consequences of this fact on the spectrum.

In the context of effective string theory, we have also reached some
conclusions about field redefinitions. We have shown that our results
on the spectrum in section \ref{sec_high} can indeed be justified in a
quantum-mechanical sense, and have motivated our position that it is
wise to avoid field-redefinitions without a careful consideration of
the transformation of the measure.  We introduced the idea of
$X$-uniformity in section \ref{fluctyxuni} and given some motivation
for its application; still, this principle also seems somewhat
restrictive and we subsequently abandoned it for a time in section
\ref{dps} in order to consider dropping the PS term itself.


\acknowledgments{
We would like to thank Pavel Boyko and Olexey Kovalenko for their valuable 
discussions and comments during the beginning of this work.
NDH wishes to thank the Hayama Center for advanced Studies for its hospitality.
}


\end{document}